\documentstyle[twocolumn,floats,psfig,aps,prb]{revtex}

\begin{document}

%\draft
\twocolumn[\hsize\textwidth\columnwidth\hsize\csname@twocolumnfalse\endcsname%
\preprint{Submitted to \prb, Rapid Communications}

\title{Non-Kondo zero bias anomaly in electronic transport through an
ultra-small Si quantum dot}

\author{L.~P. Rokhinson, L.~J. Guo, S.~Y. Chou and D.~C. Tsui}

\address{Department of Electrical Engineering, Princeton University,
Princeton, NJ 08544}

\date{Submitted to \prb, Rapid Communications on July 28, 1999}

\maketitle

\begin{abstract}
We have studied low-temperature single electron transport through
ultra-small Si quantum dots. We find that at low temperatures Coulomb
blockade is partially lifted at certain gate voltages. Furthermore, we
observed an enhancement of differential conductance at zero bias.
The magnetic field dependence of this zero bias anomaly is very
different from the one reported in GaAs quantum dots, inconsistent
with predictions for the Kondo effect.
\end{abstract}
\pacs{\\PACS numbers: 73.23.Hk, 85.30.Vw, 85.30.Wx, 75.20.Hr}

\vskip2pc]

Quantum dots (QDs) formed in GaAs/AlGaAs heterostructures have been
used as model
systems to study transport in the Coulomb blockade
regime\cite{meso-book97}.  Following advances in nanolithography QD
size was reduced down to the level where quantum effects start to play
a role. As the size become smaller collective phenomena, such as the Kondo
effect, were recently reported\cite{goldhaber98a,cronenwett98}.
To further reduce the QD size for exploring new phenomena, one must
abandon the conventional approach of using field--induced barriers.
Recently, Si dots with confinement provided by the sharp Si-SiO$_2$
interface have been realized.  These dots can be fabricated so small
that the single-electron transistor (SET) can operate at room
temperature\cite{zhuang98,ishikuro96}.  Although an increase of the
operating temperature of SETs was the primary driving force behind
the development of the Si quantum dot technology, there were some
limited studies of electron transport at low temperatures, which
provided information about energy spectrum in these
structures\cite{ishikuro97} and probed first-order quantum corrections
to the conductivity in the Coulomb blockade regime\cite{matsuoka95}.

In this paper we report low temperature electron transport in
ultra--small Si quantum dots.  We found that at certain range of gate
voltages Coulomb blockade is lifted at $T<1$ K and differential
conductance at zero source-drain bias increases as the temperature is
lowered.  Although it is appealing to attribute the enhancement to
the Kondo effect, we find that the magnetic field dependence of this
zero bias anomaly is inconsistent with such an interpretation.

We have investigated transport in quantum dot samples which are
metal--oxide--semiconductor field effect transistors (MOSFETs) with a
Si dot connected to the source and drain leads through tunneling
barriers. The dot is surrounded by 40-50 nm of SiO$_2$ and wrapped by
a poly-Si gate (fabrication details can be found in
Ref.~\onlinecite{leobandung95}). The gate is also extended over the
tunneling barriers and parts of the leads, adjacent to the
dot. Outside the gate, the source and drain are $n$-type.  An
inversion layer is formed at the Si-SiO$_2$ interface by applying a
voltage to the poly-Si gate.  Unlike GaAs dots, there are no separate
gates to control the coupling between the dot and the source/drain.
In fact, the coupling is a function of the applied gate voltage
$V_g$. We studied more than 30 samples which show Coulomb blockade
above 10 K. However, at low temperatures ($T<4$ K) and low
source--drain bias ($V_b<100$ $\mu$V) the conducting channel under the
gate breaks apart and the samples have electrical characteristics of
multiply--connected dots. Wide sweeps of the gate voltage are
accompanied by sudden switching, which could be due to
charging/discharging of some traps in the oxide.  If we restrict the
sweeps to $<1$ V, we can obtain reproducible results for several days.

\begin{figure}[t]
%\special{isoscale g-vg.wmf, 3.25in 3.25in}
%\vspace{3.3in}
\vspace{-0.5in}
\psfig{file=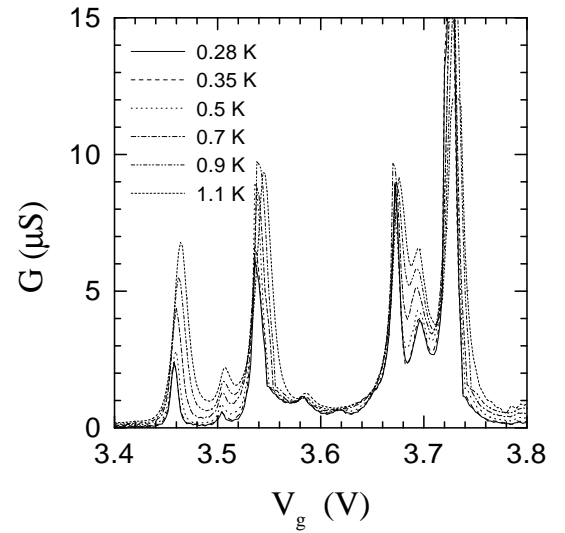,width=4.4in}
\vspace{-1.7in}
\caption{Differential conductance $G$ in the quantum dot as a function of
the gate voltage $V_g$ at six different temperatures. The $G$ is
measured with 10 $\mu$V ac source--drain bias at a frequency 7.7 Hz,
zero dc bias and $B=0$.}
\label{g-vg}
\end{figure}

In Fig.~\ref{g-vg} the differential conductance $G$ is plotted as a
function of $V_g$ at a source-drain bias $V_b=0$ for six different
temperatures from one of the samples.  From the device geometry the
dot--gate capacitance is estimated to be 1-2 aF. We attribute large
peaks at $V_g=3.46$, 3.54, 3.67 and 3.73 V to the main
lithographically defined quantum dot. From the analysis of $G$
vs. $V_b$ and $V_g$ data we estimate gate voltage -- to -- single
particle energy conversion coefficient $\alpha\approx 8.5$
mV/meV.

\begin{figure}[t]
%\special{isoscale g-t-b.wmf, 6.5in 4.1in}
%\vspace{4.1in}
\vspace{-0.9in}
\psfig{file=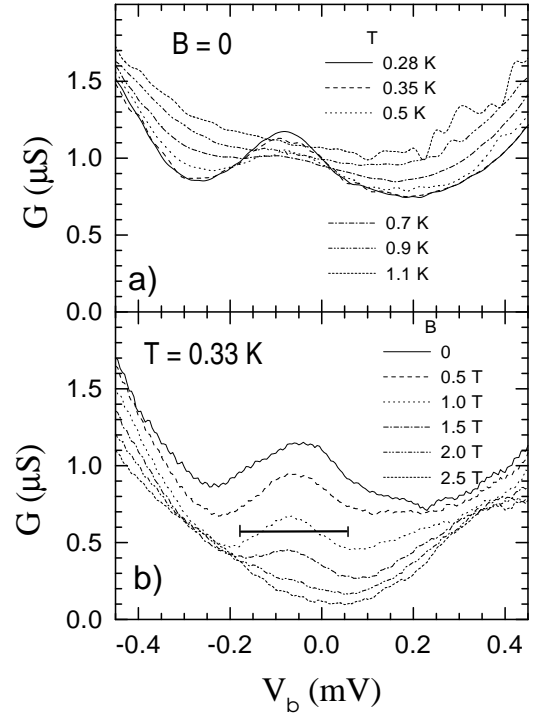,width=4.4in}
\vspace{-0.7in}
\caption{Differential conductance $G$ as a function of a dc
source-drain bias $V_b$ at different a) temperatures and b) magnetic
fields at a fixed gate voltage $V_g=3.57$ V. The a) and b) data sets
were taken one day apart that resulted in a small difference between
$T=0.35$ K data in a) and $B=0$ data in b). The bar in b) indicates
twice the Zeeman energy $2 E_Z=2 g \mu_B B=0.23$ meV ($g=2$) at $B=1$
T.}
\label{g-t-b}
\end{figure}

The $G$ in the valleys between most of the peaks is thermally
activated and vanishes rapidly at low $T$. However, in some valleys
(for example between peaks at 3.54 and 3.67 V) the $G$ is almost
$T$-independent.  Remarkably, in such valleys the $G$ vs. $V_b$ data
reveals a maximum close to the zero bias in the entire $V_g$ range of
the valley.  This is in striking contrast to the broad minimum around
$V_b=0$ observed in the neighboring Coulomb blockade valleys.  In
Fig.~\ref{g-t-b}a we plot a representative $G$ vs. $V_b$ curve
measured at $V_g=3.57$ V.  The peak at $V_b=-0.08$ mV has a weak
dependence on the $V_g$: it shifts from $V_b=-0.06$ mV at $V_g=3.55$ V
to $V_b=-0.21$ mV at $V_g=3.66$ V.  Similar results which show
slightly off--zero bias peaks that shift as a function of the $V_g$
have been reported in GaAs quantum dots\cite{schmid98,simmel99}.  In
the case of the Kondo regime, the shift of the peak from $V_b=0$ can
be qualitatively explained by the energy-- and $V_g$-- dependent
coupling of the dot to the leads.  The maximum at $V_b=-0.08$ mV
vanishes at $T>1$ K and becomes a broad minimum.

This zero bias anomaly is sensitive to an external magnetic field $B$.
At $B>2$ T, applied parallel to the conducting channel, the enhanced
conductivity is suppressed and Coulomb blockade is restored in the
entire range $3.55$ V $<V_g<3.66$ V.  At $B>2$ T the conductance shows
a broad minima near $V_b=0$.  In Fig.~\ref{g-t-b}b we plotted $G$
vs. $V_b$ at $V_g=3.57$ for different $B$. There is no apparent
dependence of the peak position on $B$, while the peak magnitude
decreases as $B$ is increased.  The peak is completely suppressed by
$B\approx2$ T.

There are striking differences between the Kondo effect reported in
GaAs dots and the zero bias anomaly in our data.  One of the
signatures of the Kondo effect is that at $B>0$ the zero bias peak in
$G$ is split into two peaks separated by twice the Zeeman energy,
$\Delta V_b=2 E_Z/e$. Such a splitting was reported in GaAs quantum
dots\cite{goldhaber98a,cronenwett98,schmid98} as well as in metallic
grains\cite{ralph94}. $E_Z=g\mu_B B=0.12$ meV at $B=1$ T (assuming
$g=2$ in Si) and the splitting is expected to be $\Delta V_b= 0.23$ mV
(indicated by a bar in Fig.~\ref{g-t-b}b).  The width of the zero bias
peak in our data is $\approx0.15$ mV at $B=1$ T, half the expected
Kondo splitting. However, we have seen no splitting of the zero bias
maximum as a function of $B$ in our data up to $B=2$ T, the highest
field at which the maximum is still observed.  Also, the position of
this maximum is not effected by magnetic field.

An underlying physics for the Kondo effect requires the highest
occupied level in the dot to be at least doubly degenerate. Adding an
extra electron to the dot costs just a bare charging energy
$U_c=e^2/2C$, where $C$ is the total capacitance from the dot to the
gate and leads. Adding a second electron should cost $U_c + \Delta E$,
where $\Delta E$ is due to the size quantization in the dot (or it can
be the same $U_c$ if the level is more than two--fold degenerate).
Thus, the Kondo effect is expected to be observed in the narrower
valley between two adjacent charge--degenerate peaks which are
separated by $\Delta V_g=U_c/\alpha$, while neighboring valleys are
expected to be wider with gate voltage separation of $\Delta
V_g=(U_c+\Delta E)/\alpha$.  However, we observed zero bias anomaly in
the widest valley with $\Delta V_g=170$ mV, while the neighboring
valleys with widths 80 and 60 mV have no zero bias anomalies at
$B=0$. There is also a characteristic shift of the charge--degeneracy
peaks as a function of temperature in the Kondo
regime\cite{goldhaber98}.  As temperature decreases, off-resonant
conductance is enhanced which results in the shift of the
charge--degeneracy peaks toward each other.  Instead, we observed that
the peak at $V_g=3.54$ V shifts to the lower gate voltages as the
temperature is decreased, while the position of the peak at $V_g=3.67$
V is almost temperature independent.

\begin{figure}[t]
%\special{isoscale energ.wmf, 3.25in 3.25in}
%\vspace{3.3in}
\vspace{-0.5in}
\psfig{file=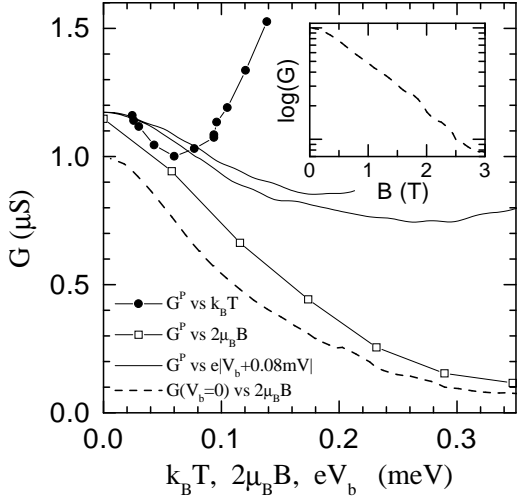,width=4.4in}
\vspace{-1.7in}
\caption{Peak differential conductance $G^P$ at $V_b=-0.08$ mV is
plotted as a function of temperature $k_B T$ at $B=0$ ($\bullet$),
magnetic field $2\mu_B B$ at $T=0.3$ K ($\Box$). The two solid curves
are $G^P$ vs. bias $e|V_b-V_b^P|$ at $B=0$ and $T=0.28$ K, where
$V_b^P=-0.08$ mV is the peak position. Dashed line is $G$ vs. $B$ at
$V_b=0$ and $T=0.3$ K. Inset: $G$ at $V_b=0$ falls almost
exponentially as a function of $B$.}
\label{energ}
\end{figure}

Dependence of the conductance of the zero bias peak $G^P$ on $T$, $B$
and $V_b$ is shown in Fig.~\ref{energ}.  The temperature range
$0.3<T<1$ K, where zero bias anomaly is observed, is not sufficient to
extract the functional dependence $G^P(T)$ with certainty, although
it is close to being logarithmic. The zero bias peak is superimposed on
a parabolic $V_b$--dependent background, thus we cannot unambiguously
conclude what is the functional dependence of $G^P$ on the bias
voltage.  In contrast, $G^P$ is a strong function of the magnetic
field.  As shown in the inset in Fig.~\ref{energ}, magnetic field
exponentially suppresses the conductance by more than an order of
magnitude.  Note, that one expects a weak logarithmic suppression of
$G$ by magnetic field at $V_b=0$ in the Kondo regime.

\begin{figure}[t]
%\special{isoscale others.wmf, 3.25in 3.5in}
%\vspace{3.5in}
\vspace{-0.5in}
\psfig{file=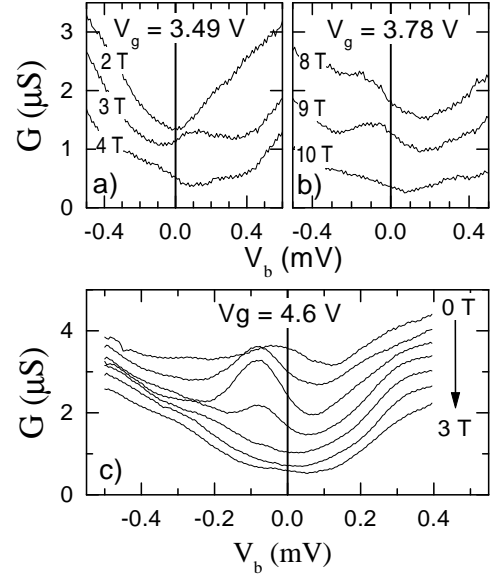,width=4.4in}
\vspace{-1.4in}
\caption{Differential conductance $G$ as a function of dc bias $V_b$
in three different Coulomb blockade valleys. In these valleys peaks in
$G$ are observed at $B\neq0$. Curves are offset by 0.5 $\mu$S in a)
and b) and by 0.25 $\mu$S in c), except for the bottom curves in each plot.}
\label{others}
\end{figure}

Another striking result is that in some Coulomb blockade valleys the
zero bias anomaly appear only at non-zero magnetic field. These
valleys may group around the valley where the zero bias anomaly is
observed at $B=0$.  For example, at $B=0$ we observe a peak in $G$ at
$V_b\approx0$ in the valley 3.54 V $<V_g<$ 3.67 V (Fig.~\ref{g-t-b}),
while there are broad minima at $V_b=0$ in the neighboring valleys
(3.46 V $<V_g<$ 3.54 V and 3.67 V $<V_g<$ 3.72 V). The zero bias
anomaly peak at 3.55 V $<V_g<$ 3.66 V is destroyed by $B\approx2$ T
and $G$ has a broad minimum centered at $V_b=0$ at higher magnetic
field. However, at $B=3$ T $G$ has a maximum in the valley 3.46 V
$<V_g<$ 3.54 V. In Fig.~\ref{others}a the $V_b$--dependence of $G$ is
shown in the center of that valley at $V_g=3.49$ V. While there is a
minimum around $V_b=0$ at $B<2$ T and $B>4$ T, there is a pronounced
peak at $V_b= 0.1$ mV at $B=3$ T. In the neighboring valley (at
$V_g=3.707$ V) we observed a peak at $V_b=-0.3$ mV at $B=5$ T. At yet
higher $B=9$ T there is a maximum at $V_b= -0.1$ mV at $V_g=3.78$ V,
as shown in Fig.~\ref{others}b.  These maxima are observed over a
limited $B$ range of $\Delta B\approx 1$ T.

In some Coulomb blockade valleys zero bias anomaly is observed at
$B=0$, although the strongest zero bias peak is found at $B>0$. As
shown in Fig.~\ref{others}c, at $V_g=4.6$ V the strongest zero bias
peak is at $B=0.6$ T; the peak becomes a broad minimum at $B>2.5$~T.
There are no zero bias anomalies developed in the adjacent valleys in
the experimental range of $0<B<10$ T.

To summarize our findings, we observed a suppression of the Coulomb
blockade and an enhancement of the differential conductance at low
temperatures at certain gate voltages.  This anomaly is destroyed by
i) raising the temperature, ii) increasing the bias, or iii) applying
a magnetic field.  Unlike in the Kondo effect, the zero bias anomaly
in our experiment is not split by the magnetic field but, instead, the
magnetic field suppresses it exponentially. Also, at certain gate
voltages we observed a zero bias anomaly at $B>0$.  Our data cannot be
understood within the framework of the theories of the Kondo effect.

Authors gratefully acknowledge discussions with Ned Wingreen. The work
was supported by ARO, ONR and DARPA.


\begin{references}

\bibitem{meso-book97}
L.~P. Kouwenhoven, C.~M. Marcus, P.~L. McEuen, S. Tarucha, R.~M. Westervelt,
  and N.~S. Wingreen,  in {\em Mesoscopic Electron Transport}, Vol.~345 of {\em
  NATO ASI series}, edited by L.~L. Sohn, L.~P. Kouwenhoven, and G.Sch\"{o}n
  (Kluwer Academic Publishers, Boston, 1997).

\bibitem{goldhaber98a}
D. Goldhaber-Gordon, H. Shtrikman, D. Mahalu, D. Abusch-Magler, U. Meirav, and
  M.~A. Kastner, Nature {\bf 391},  156  (1998).

\bibitem{cronenwett98}
S.~M. Cronenwett, T.~H. Oosterkamp, and L.~P. Kouwenhoven, Science {\bf 281},
  540  (1998).

\bibitem{zhuang98}
L. Zhuang, L. Guo, and S.~Y. Chou, \apl {\bf 72},  1205  (1998).

\bibitem{ishikuro96}
H. Ishikuro, T. Fujii, T. Saraya, G. Hashiguchi, T. Hiramoto, and T. Ikoma,
  \apl {\bf 68},  3585  (1996).

\bibitem{ishikuro97}
H. Ishikuro and T. Hiramoto, \apl {\bf 71},  3691  (1997).

\bibitem{matsuoka95}
H. Matsuoka and S. Kimura, \apl {\bf 66},  613  (1995).

\bibitem{leobandung95}
E. Leobandung, L. Guo, Y. Wang, and S.~Y. Chou, \apl {\bf 67},  938  (1995).

\bibitem{schmid98}
J. Schmid, J. Weis, K. Eberl, and K. v.~Klitzing, Physica B {\bf 256},  182
  (1998).

\bibitem{simmel99}
F. Simmel, R.~H. Blick, J.~P. Kotthaus, W. Wegscheider, and M. Bichler,
  cond-mat/9812153 (unpublished).

\bibitem{ralph94}
D.~C. Ralph and R.~A. Buhrman, \prl {\bf 72},  3401  (1994).

\bibitem{goldhaber98}
D. Goldhaber-Gordon, J. G\"{o}res, M.~A. Kastner, H. Shtrikman, D. Mahalu, and
  U. Meirav, \prl {\bf 81},  5225  (1998).

\end{references}
\end{document}